\documentclass[twocolumn,showpacs,preprintnumbers,amsmath,amssymb,nofootinbib]{revtex4}

\usepackage{graphicx}
\usepackage{dcolumn}
\usepackage{bm}
\usepackage{epsfig}
\usepackage{amssymb}
\usepackage{amsmath}

\def\fun#1#2{\lower3.6pt\vbox{\baselineskip0pt\lineskip.9pt
  \ialign{$\mathsurround=0pt#1\hfil##\hfil$\crcr#2\crcr\sim\crcr}}}

\begin{document}

\title{The inflationary gravitational-wave background and measurements of the scalar spectral index}

\author{Tristan L. Smith$^1$, Marc Kamionkowski$^1$, and Asantha
Cooray$^{2}$}
\affiliation{$^1$California Institute of Technology, Mail Code
130-33, Pasadena, CA 91125}
\affiliation{$^2$Department of Physics and Astronomy, University
of California, Irvine, Irvine, CA 92697}

\date{\today}
   
\begin{abstract}
Inflation predicts a stochastic background of
gravitational waves over a broad range of frequencies, from
those accessible with cosmic microwave background (CMB)
measurements, to those accessible directly with
gravitational-wave detectors, like NASA's Big-Bang Observer
(BBO), currently under study.  
In a previous
paper [Phys.~Rev.~{\bf D73}, 023504 (2006)] we connected CMB constraints 
to the amplitude
and tensor spectral index of 
the inflationary gravitational-wave background (IGWB) at BBO
frequencies for four classes of models of inflation by directly solving the inflationary
equations of motion.
Here we extend that analysis by including results obtained in the
WMAP third-year data release as well as by considering two additional classes of inflationary 
models.
As often noted in the literature, the recent indication that the primordial density power-spectrum has a red spectral index implies (with some caveats) that 
the amplitude of the IGWB may be large enough to be observable in the CMB polarization.  Here we also explore the implications
for the direct detection of the IGWB.  
\end{abstract}


\maketitle

\section{Introdtuction}

With the advent of precise cosmic microwave background (CMB)
measurements
\cite{Kamionkowski:1999qc, deBernardis:2000gy, Miller:1999qz,
Hanany:2000qf, Halverson:2001yy, Mason:2002tm, Benoit:2002mm,
Goldstein:2002gf, Spergel:2003cb}, inflationary cosmology has now
become an empirical science.  The inflationary paradigm was proposed nearly three decades ago in order
to address several theoretical deficiencies with the standard cosmological scenario
\cite{Guth81,Albrecht:1982wi,Linde:1981mu}.
  The
concordance of the cosmological measurements with the inflationary
expectations of a flat universe and a nearly scale-invariant
spectrum of primordial density perturbations
\cite{Guth82,Bardeen:1983qw,Hawking:1982cz,Linde:1982uu} is at least
suggestive and warrants further tests of inflation.  One of the most
unique and exciting
predictions of inflation yet to be tested is the existence of a stochastic
gravitational-wave background with a nearly scale-invariant
spectrum \cite{Starobinsky:1979ty,Abbott84,Starobinskii,Rubakov:1982df,Fabbri:1983us, 
Allen:1987bk, Sahni:1990tx}.
Detection of the CMB B-mode polarization pattern induced by
inflationary gravitational waves of wavelengths
comparable to the horizon has become a goal of next-generation
CMB experiments
\cite{Kamionkowski:1996ks, Kamionkowski:1996zd,
Zaldarriaga:1996xe, Seljak:1996gy,Cabella:2004mk}.

In a previous paper \cite{Smith:2005mm} we surveyed four classes of inflationary models to investigate how 
CMB constraints to the inflationary cosmology translates into allowed regions in 
the plane spanned by the IGWB amplitude and tilt at frequencies that correspond to  
direct detection satellite experiments.  We concentrated on 
determining whether the IGWB would be observable in future space-based gravitational-wave observatories such as NASA's Big Bang Observer (BBO) \cite{BBO}, currently under study \cite{phinney_comm,Seto:2005qy}.  

Recent results from the WMAP team \cite{Spergel:2006hy} (published after our initial paper) indicate that the slope of the
scalar perturbations, $n_s$, is different from scale invariant with a best fit value at $n_s \approx 0.95$ (for a zero tensor contribution).  
Although some groups have challenged the exact statistical significance of this result (see, e.g., Refs.~\cite{Eriksen:2006xr, Kristiansen:2006ec}) 
 the conclusion that $n_s<1$, if upheld by future observations, may have important implications 
for the IGWB.  In particular, as argued in Refs.~\cite{Boyle:2005ug,Pagano:2007st}, the confirmation of a spectral index less than one may indicate (with caveats that we explore below) that the effects of the IGWB on the CMB polarization pattern will be large enough to be detectable with future missions \cite{spider, planck}.  In this paper we discuss how the curvature of the inflaton potential determines, to a large extent, whether or not finding $n_s <1$ implies a large IGWB amplitude.  The same reasoning applies to the chances of directly observing the relic IGWB today with future space-based gravitational-wave observatories.  

In order to assess how the most recent data impacts our future ability to directly observe the IGWB we have re-analyzed how the most recent CMB constraints translate into predictions for the IGWB at direct detection scales for the original four classes of single field inflationary models analyzed in Ref.~\cite{Smith:2005mm} as well as for two new classes of models.  The range of models analyzed in this paper allows us to understand the general implications for the detection of the IGWB when the primordial density slope is significantly different from unity. 

This paper is organized as follows.  In Section II we present an abbreviated summary of 
the method used in Ref.~\cite{Smith:2005mm} to connect CMB constraints to the inflationary parameters with the parameters probed by gravitational-wave observatories.  We refer the reader to that reference for details.  In Section III, we review the
arguments that show that a scale-dependent density-perturbation spectrum
implies a large value for the amplitude of the IGWB.  In Section IV we present 
our results for the six classes of models of inflation considered in this paper.  
 In
Section V, we summarize and make some concluding remarks.
Throughout this paper we shall make use of quantities that are defined in Ref.~\cite{Smith:2005mm}.  We refer the reader to that paper for their definitions and further discussion. 

\section{Range of inflationary parameters}

For the models that we consider we allow the number, $N$, of $e$-folds of inflation after the current horizon exited the horizon during inflation to range between $47 \leqslant N \leqslant 62$.  This corresponds to allowing the reheat temperature to range between $10^{16}$ GeV and 1 MeV.  We note that the frequencies accessible to space-based gravitational-wave observatories probe epochs ($\sim 0.1-1$ Hz) which exited the inflationary horizon about 35 $e$-folds after the current horizon.  We furthermore fix the amplitude of the primordial power spectrum to be $P_s(k_0) = 2.45 \times 10^{-9}$ \cite{Seljak04} and require the running of the scalar spectral index to be $|\alpha_s| \lesssim 0.044$ at a pivot wavenumber $k_0 = 0.05 \
\mathrm{Mpc}^{-1}$ \cite{Seljak:2006bg}.  Any error in a measurement of the amplitude of the primordial power spectrum will cause a slight additional broadening beyond
that due to the allowed range in $N$.  

We note here that the gravitational-wave transfer function given in Eq.~(26) in Ref.~\cite{Smith:2005mm} is wrong by a factor of 2 as a result of not properly taking into account the double sided Fourier transform used to define the gravitational-wave power-spectrum.  
The transfer function can be written in the form \cite{Easther:2006gt}
\begin{equation}
\Omega_{\mathrm{gw}}h^2 = \frac{32}{9} \Omega_r h^2 \left(\frac{g_*(T_0)}{g_*(T_k)}\right)^{1/3} \frac{V}{m_{\mathrm{pl}}^4}, 
\end{equation}
where $\Omega_r \equiv \rho_r^0/\rho_c^0$,
\begin{eqnarray}
\rho_r^0 &=& \frac{\pi^2}{30} T_0^4 g_*(T_0), \\
\rho_c^0 &=& \frac{3 H_0^2 m_{pl}^2}{8 \pi}, 
\end{eqnarray}
\begin{figure}
\centerline{\epsfig{file=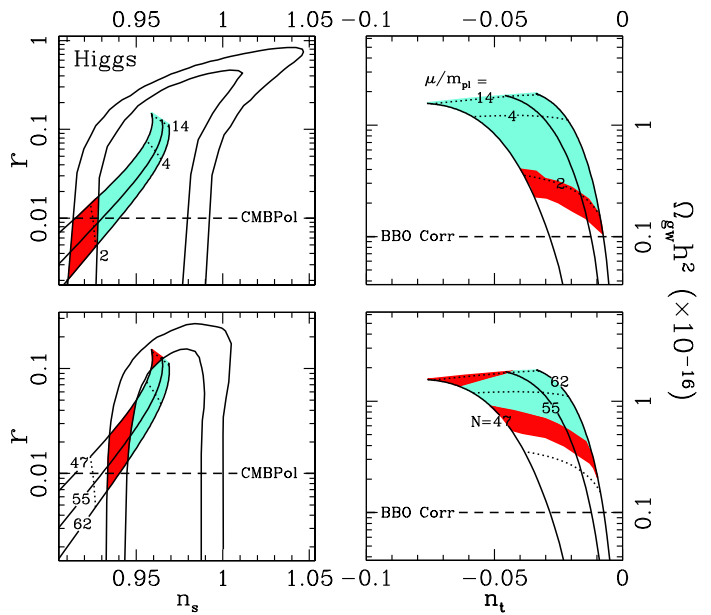,height=19pc,angle=0}}
\caption{Results for the Higgs potential.  The upper left panel shows the CMB constraints (68\% and 95\% contours) imposed by just considering the WMAP third year data \cite{Spergel:2006hy}; the lower left panel shows the CMB constraints (68\% and 95\% contours) imposed by considering a suite of data including measurements of galaxy clustering and Lyman-alpha forest constraints as described in Ref.~\cite{Seljak:2006bg}.  The dashed lines on the left-hand panels indicate $r=0.01$ roughly the limit for CMBPol \cite{Bock:2006yf}.  The panels on the right show the corresponding predictions for the IGWB given the CMB constraints.  The dashed lines on the right-hand panels indicate the sensitivity of the second generation BBO interferometer known as `BBO correlated' \cite{phinney_comm,Seto:2005qy}.  The solid black lines indicate directions of constant number of $e$-folds of inflation and the dotted black lines indicate directions of constant minimum field value, $\mu$.}
\end{figure}
$V$ is the inflaton potential, $H_0$ is the Hubble parameter today, $T_0 = 2.73$ K \cite{Mather:1993ij} is the photon temperature today, and $m_{\mathrm{pl}} = 1.2 \times 10^{19}$ GeV is the Planck mass. 
 We have supposed that the universe was radiation dominated when length scales corresponding to the direct detection of the IGWB re-entered the horizon ($T_k \sim 10^7$ GeV).  
 Photons plus three species of massless neutrinos [$g_*(T_0)=3.91$] give
 \begin{equation}
 \Omega_{\mathrm{gw}}h^2 = A_{\mathrm{gw}} \frac{V}{m_{\mathrm{pl}}^4},
 \end{equation}
 where $A_{\mathrm{gw}} = 6.08 \times 10^{-5} g_{100}^{-1/3}$ and $g_{100} \equiv g_{*}(T_k)/100$.
 
 \section{The connection between $n_s < 1$ and the IGWB amplitude}
 
Many authors \cite{Boyle:2005ug,Pagano:2007st} have noted that when $n_s \neq 1$ then the amplitude of the IGWB is, generically, significant.  The argument for this conclusion is made by looking
at the expression for $n_s$ in terms of the slow-roll parameters $\epsilon$ and $\eta$, 
\begin{equation}
1-n_s = 6 \epsilon - 2 \eta,
\label{ns}
\end{equation}
and the tensor-to-scalar ratio
\begin{equation}
r \equiv \frac{A_t}{A_s} = 16 \epsilon.
\label{r}
\end{equation}
The slow-roll parameters are given in terms of the inflaton potential by
\begin{eqnarray}
     \epsilon &\equiv& \frac{m_{\mathrm{pl}}^2}{16 \pi}
     \left(\frac{V^{\prime}}{V}\right)^2, \\
     \eta &\equiv&  \frac{m_{\mathrm{pl}}^2}{8 \pi}
     \frac{V^{\prime \prime}}{V}.
\end{eqnarray}

\pagebreak
In order to infer a value for $r$ given the indication that $1-n_s \approx 0.05$\footnote{When making this argument we ignore the fact that current analyses indicating $n_s <1$ fix $r=0$.  The full analysis presented in this paper does not fix $r$.} one has to suppose some `natural' relationship between $\epsilon$ and $\eta$.   Many authors have supposed that $\epsilon \gtrsim \eta$ so that `at worst' we have $\mathcal{O}(\epsilon) \approx \mathcal{O}(\eta)$.  We can then conclude that $1-n_s \approx 0.05$ implies that $r \approx \mathcal{O}(0.1)$.  Such a value for the tensor-to-scalar ratio is easily accessible to future CMB experiments as well as to space-based gravitational-wave observatories.  However, as we shall now show, there is are caveats when making the above argument.  

Taylor expanding the potential in terms of these slow-roll parameters evaluated at some value of the inflaton field, $\phi_*$, we obtain \cite{Peiris:2006ug}
\begin{eqnarray}
V(\phi) &\approx& \frac{m_{\mathrm{pl}}^2 H_*^2}{8 \pi} \bigg\{(3-\epsilon_*) \label{pot}\\
&&+ 12 \sqrt{\pi\epsilon_*} \frac{\delta \phi}{m_{\mathrm{pl}}} \nonumber\\
&&+ 2 \pi[3 \sqrt{\epsilon_*/\pi}  + 6(\eta_*-\epsilon_*)]\left(\frac{\delta \phi}{m_{\mathrm{pl}}}\right)^2\bigg\}, \nonumber
\end{eqnarray} 
where $\delta \phi \equiv \phi-\phi_*$.
If we suppose that $\epsilon_* \ll \eta_*$ then we can see that the potential is well approximated by a quadratic function with the coefficient $(3/2) H_*^2 m_{\mathrm{pl}}^2\eta_*$.  Looking at Eq.~(\ref{ns}) the fact that $1-n_s >0$ implies $\eta_* < 0$ in this case--- i.e., the curvature of the potential must be negative. 
 \begin{figure}
\centerline{\epsfig{file=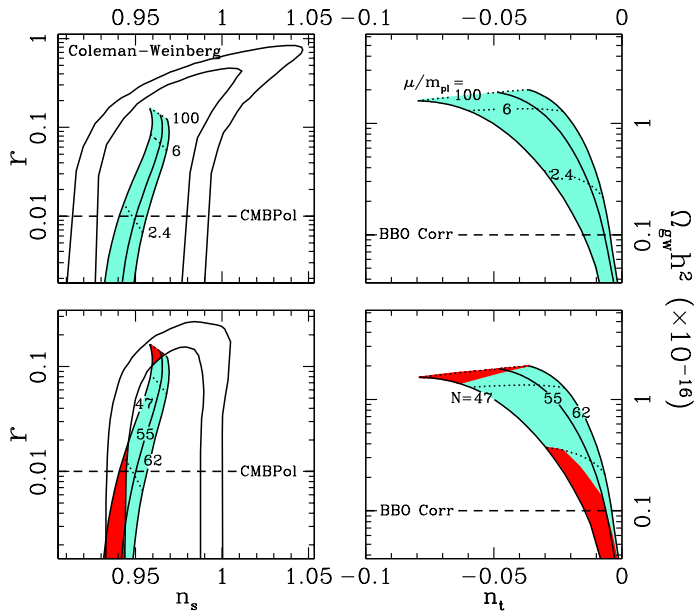,height=19pc,angle=0}}
\caption{Same as Fig.~1 but for the Coleman-Weinberg potential.}
\end{figure}

For the case where $\eta <0$ at some point, in order for inflation to lead to oscillations in $\phi$ resulting in the reheating of the universe, $\eta$ must change sign so that the field evolves into a minimum of the potential.  In Ref.~\cite{Boyle:2005ug} the fact that $\eta$ must change sign was used to indicate fine tuning.  However, as Ref.~\cite{Boyle:2005ug} points out, there are several scalar-field potentials that have this property as a result of particular symmetries (such as the Higgs potential) so that, in some sense, their `fine tuning' is justified.  As we shall see, it is exactly these potentials that allow for $n_s < 1$ and $r \ll 1-n_s $. 
 
 \section{CMB constraints and direct detection amplitudes}
 
 In this section we present the results of our analysis.  We divide our presentation into two parts.  First we discuss potentials for which $n_s < 1$ and $r \ll 1-n_s$.  These potentials all share the property that inflation starts at a flat section of the potential and then rolls over a negatively curved region to a global minimum.  The second set of potentials share the property that the inflaton field starts on a positively curved region of the potential and inflation ends as the field enters a global minimum or as a second field becomes dynamically important.  We present the functional forms of the various potentials considered here in Table I. 
 
 In our analysis we use the CMB constraints derived from considering the WMAP third-year data release (WMAP3, Ref.~\cite{Spergel:2006hy}) and those derived by considering a suite of CMB observations (including WMAP3) as well as measurements of the linear matter power-spectrum coming from the Sloan Digitial Sky Survey (SDSS) and the Lyman-alpha forest, measurements of the baryon acoustic peak from SDSS, and measurements of supernovae luminosity distances (WMAP3+, Ref.~\cite{Seljak:2006bg}).  We note that the analysis in Ref.~\cite{Spergel:2006hy} fixes the pivot wavenumber at $k_0 = 0.002\ \mathrm{Mpc}^{-1}$ whereas the analysis in Ref.~\cite{Seljak:2006bg} uses $k_0 = 0.05\ \mathrm{Mpc}^{-1}$.  As pointed out in Ref.~\cite{Cortes:2007ak}, the overall constraints do not depend on the choice of $k_0$, but the value of $k_0$ may change the shape of contours for marginalized constraints (such as contours in the $n_s-r$ plane).  The CMB constraints used in this paper fix the running of the scalar spectral index to zero and therefore, according to the analysis in Ref.~\cite{Cortes:2007ak}, are only slightly affected by the choice of pivot wavenumber.  
  
 \subsection{Potentials with $n_s < 1$ and $r \ll 1-n_s$ }
 
 As one can infer from the expansion of the inflaton potential given in Eq.~(\ref{pot}) those potentials for which at some field value $\epsilon \ll \eta$ and $\eta <0$ can be approximated by $V(\phi) \approx V_0(1-V_1^2 [\delta\phi/m_{\mathrm{pl}}]^2)$.  
 \begin{table}[tb]\footnotesize
\caption{\label{tab:specs}}
\begin{center}
{\sc Potentials considered in this paper\\}
\begin{tabular}{rc} 
\\
\hline
\hline
%\tableskip\tableline\tableline\tableskip
Potential name & $V(\phi)$ \\
\hline \\
%\tableskip\tableline\tableskip
Higgs: 
& $V_0(1-[\phi/\mu]^2)^2$ \\
\\
%\tableskip\tableline\tableskip
Coleman-Weinberg: 
& $V_0\bigg\{ \left(\frac{\phi}{\mu}\right)^4 \Big(\log\left[\frac{\phi}{\mu}\right] - \frac{1}{4}\Big) + \frac{1}{4}\bigg\}$ \\
%\tableskip\tableline
\\
PNGB:
&$V_0[1-\cos(\phi/\mu)]$  \\
\\
Chaotic:
&$V_0 \left(
     \frac{\phi}{m_{\mathrm{pl}}}\right)^{p}$\\
     \\
 Power-law:
 & $V_0 e^{-p \phi/m_{\mathrm{pl}}}$ \\
 \\
 Hybrid:
 & $V_0\left[1+\left(\frac{\phi}{\mu}\right)^2 \right]$ \\
 \\
\hline
\end{tabular}
\end{center}
\end{table}
 The early evolution of the inflaton field in these models can be described as rolling off of a `cliff': inflation starts at the nearly flat region near the origin and then rolls down the negative slope (i.e., $\eta <0$).  Potentials of this form were also investigated in Refs.~\cite{Boubekeur:2005zm,Kohri:2007gq}.  Inflation ends as the field reaches a global minimum at some field value $\mu$.  If scales corresponding to the current horizon exited during the initial descent down this slope these models generically predict $n_s < 1$ and $r \ll n_s -1$. 

We analyzed three classes of models that possess this behavior: the Higgs, Coleman-Weinberg 
\cite{Linde:1982uu,Albrecht:1982wi,Shafi:1984tt,Shafi:2006cs, Kinney:1995ki}, and pseudo Nambu-Goldstone boson (PNGB) potentials \cite{Freese:1990rb,Freese:2004un,Savage:2006tr,Kinney:2007np}.  Each of these potentials is characterized by an amplitude and a field value at which the potential is minimized.  As discussed in Ref.~\cite{Smith:2005mm} the end of inflation is fixed by the requirement $\epsilon(\phi_{\mathrm{end}}) = 1$ and the amplitude of the potential is determined by the requirement that the density perturbations have an amplitude $\sim 10^{-5}$.  
Therefore, for each potential we are free to vary the number of $e$-folds of inflation after the current horizon exited the horizon during inflation as well as the field value at which the potential is minimized.  This two-dimensional freedom can be translated into a region in the plane spanned by $(r,n_s)$ for the CMB and $(n_t, \Omega_{\mathrm{gw}}h^2)$ for gravitational-wave observatories. 

For these models, the constraints obtained with WMAP3+ are more stringent than those derived from just analyzing WMAP3 data.  This is a result of the fact that the WMAP3+ constraints include an analysis of the matter power-spectrum on small length scales derived from observations of the Lyman-alpha forest.  This allows for a better constraint to the \emph{lower} bound of $n_s$ while at the same time the analysis improves the bound on $r$ by removing some of the degeneracy between $r$ and $n_s$ found in a CMB-only analysis. 
\begin{figure}
\centerline{\epsfig{file=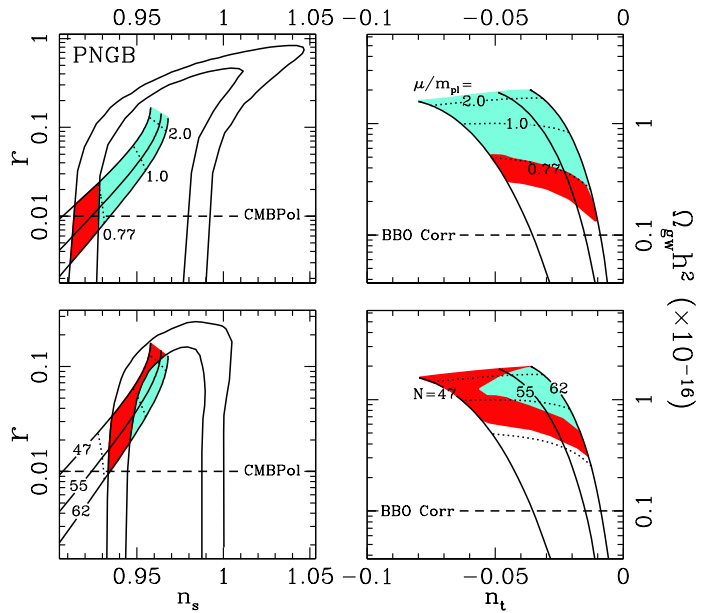,height=19pc,angle=0}}
\caption{Same as Fig.~1 but for the PNGB potential.}
\end{figure}
As we can see from Figs.~1, 2, and 3 a \emph{lower} bound to the IGWB produced by this class of models will improve as we improve the \emph{lower} limit to the scalar spectral index. 

One qualitative difference between these three classes of potentials is clear when comparing Higgs and PNGB inflation to Coleman-Weinberg inflation.  Both Higgs and PNGB inflation trace out similar regions in the $n_s-r$ plane where $\mathrm{d}r/\mathrm{d} n_s \sim 1$.  In contrast to this behavior, Coleman-Weinberg inflation traces out a region with a fairly large slope, with $r$ decreasing rapidly around $n_s \sim 0.94$.  This is a result of the fact that the Coleman-Weinberg potential remains particularly flat for a larger range in field values around the origin as compared to the Higgs or PNGB potentials.  

\subsection{Potentials with $r \sim |1-n_s|$}
 \begin{figure}
\centerline{\epsfig{file=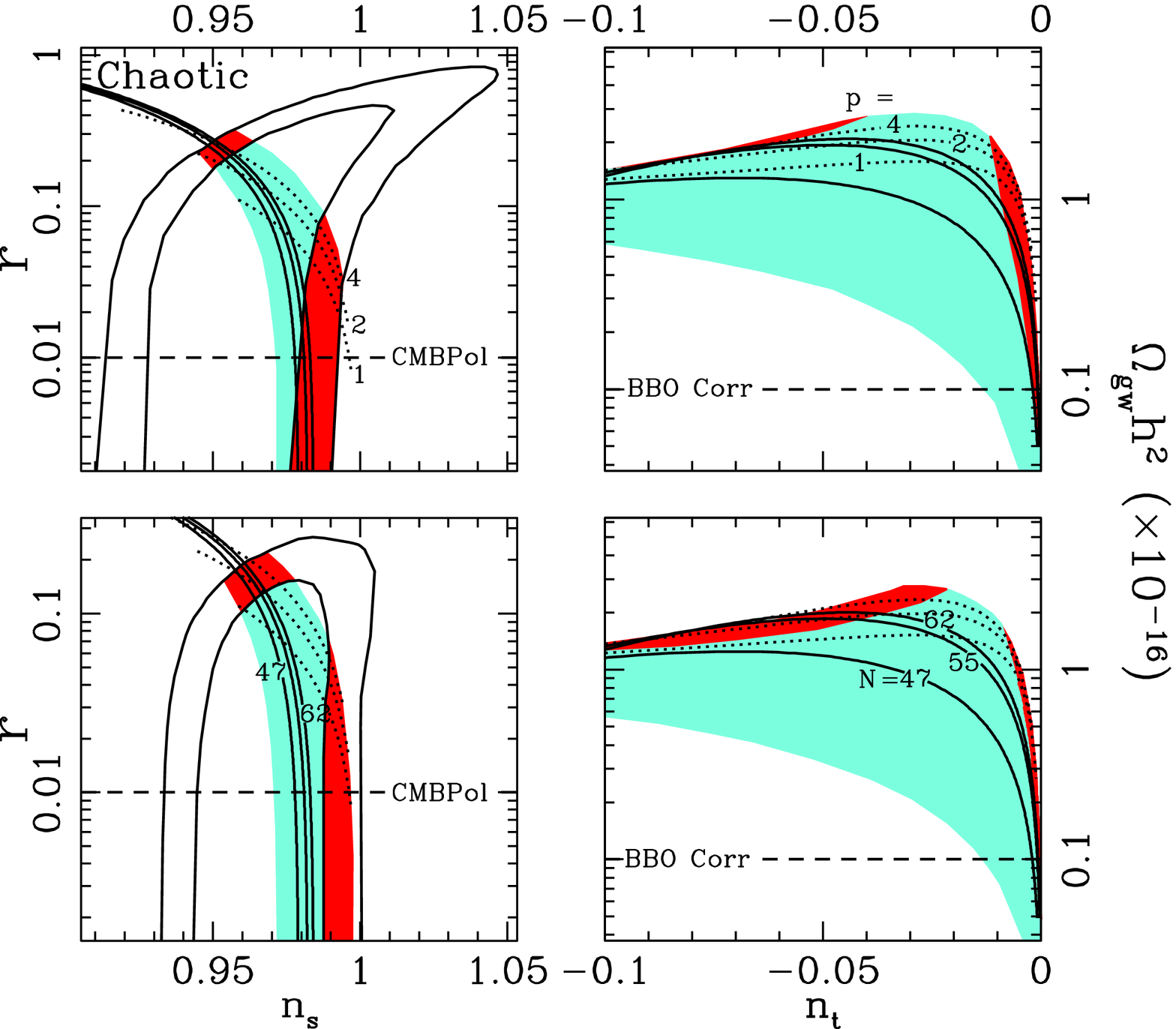,height=19pc,angle=0}}
%\end{center}
\caption{Same as Fig.~1 but for the chaotic potential.  As commented in the text, only those models for which the index $p$ is a positive even integer allow for a proper end to inflation.  For other choices of $p$ the form of the inflaton potential must change before inflation ends.  As a result, we allow for the field value $\phi_{\mathrm{CMB}}$ to be a free parameter, as discussed in the text.  In order to indicate the predictions for those models in this class that reach a proper end of inflation (i.e., where $p$ is a positive even integer), the solid black lines correspond to between 62 and 47 $e$-folds of inflation and the dotted lines indicate constant values for the index of the potential.  As is commented in Ref.~\cite{Spergel:2006hy} a massive scalar field ($p =2$) is a good fit to the data whereas a quartic potential lies outside of the 2$\sigma$ confidence region using just WMAP3 data.  This disagreement is worsened when using the WMAP3+ constraints.}
%\end{figure*}
\end{figure}

We now consider the results for three potentials (chaotic, hybrid, and power-law inflation) which respect the relation $r \sim |1-n_s|$.  These models are characterized by the property that the current horizon exited the horizon during inflation when the inflaton field sat at a point of the potential with positive curvature (i.e., $\eta > 0$).  
%\begin{figure*}
  %\begin{center}
%\leavevmode
\begin{figure}
\centerline{\epsfig{file=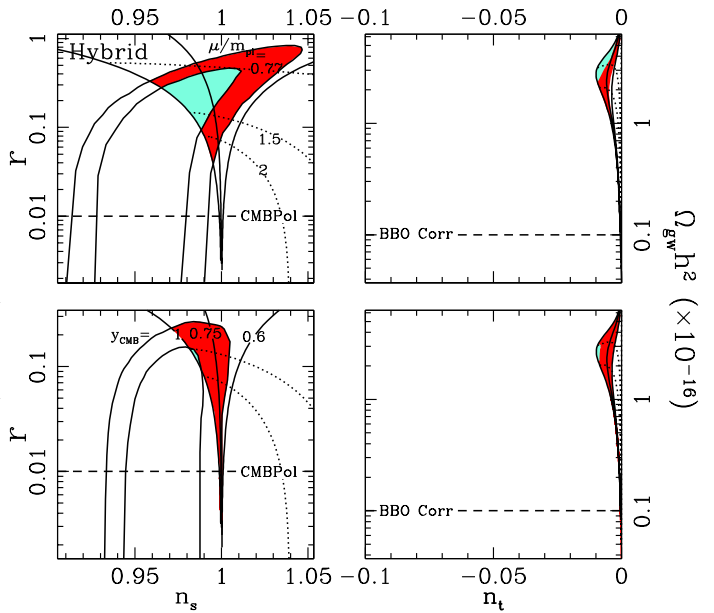,height=19pc,angle=0}}
%\end{center}
\caption{Similar to Fig.~1 but for the hybrid potential.  Unlike Fig.~1 the solid black lines
follow curves of constant $y_{\mathrm{CMB}} \equiv \phi_{\mathrm{CMB}}/\mu$.  See the text for further discussion.}
%\end{figure*}
\end{figure}
  \begin{figure}[h!]
\centerline{\epsfig{file=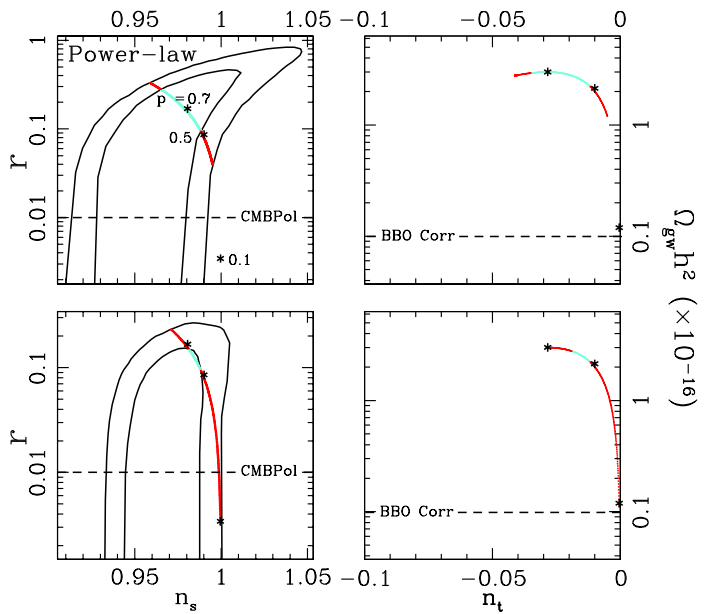,height=19pc,angle=0}}
\caption{Similar to Fig.~1 but for power-law inflation.  The stars indicate values for the power-law index, $p$.  For a fixed scalar amplitude, $P_s$, this model occupies only a line in the $n_s-r$ plane because both $\epsilon$ and $\eta$ are functions only of the index $p$ and not of the field value.}
\end{figure}

Unlike the models considered in the last subsection, the free parameter in these models does not control the value of the field when inflation ends.  In particular, both power-law inflation and hybrid inflation are thought to end through some mechanism other than the evolution of the inflaton field.  Therefore, for these models we take the value of the inflaton field corresponding to CMB observations ($\phi_{\mathrm{CMB}}$) as a free parameter (instead, as in the last section, the number, $N$, of $e$-folds of inflation after this point).  In the case of hybrid inflation, we require that inflation end before the field reaches $\phi=0$ and we only consider $\phi_{\mathrm{CMB}}/\mu \leq 1$--- for values greater than unity the dynamics is closely related to chaotic inflation with $p =2$ \cite{Copeland:1994}. 
 
A slight exception to this is chaotic inflation in which inflation ends when the inflaton field attains the value $\phi = p\ m_{\mathrm{pl}}/(4 \sqrt{\pi})$ where $p$ is the index of the monomial potential, $V \propto \phi^{p}$ (see Table I).  However, in order for inflation to end and for oscillations in the inflaton field to begin, $p$ must be even.  By considering models for which $p$ is not even we are implicitly supposing that the form of the potential changes between the field values corresponding to CMB and gravitational-wave observatory observations and the end of inflation.  In order to take this into account we allowed the field value corresponding to CMB observations, $\phi_{\mathrm{CMB}}$, to be a free parameter, only requiring that it be at least 35 $e$-folds before the field reached the value $\phi = p\ m_{\mathrm{pl}}/(4 \sqrt{\pi})$.  If we allowed for a value for $\phi_{\mathrm{CMB}}$ to be lower then this, then the form of the potential must change between field values corresponding to CMB observations and the direct observation of the IGWB.  Such a situation was explored in Ref.~\cite{Smith:2005mm} in the form of a broken scale-invariant potential.

As in the previous section, the normalization of each potential is set by the requirement that it produce the appropriate amplitude for scalar perturbations when evaluated at the field value, $\phi_{\mathrm{CMB}}$, corresponding to CMB observations.  

Besides the freedom to set $\phi_{\mathrm{CMB}}$, with the restrictions discussed before, each model has one free parameter which we also vary (see Table I).  
The model parameter plane can then be 
mapped on to the CMB plane $(n_s, r)$.  Constraints derived from CMB observations are then translated to the plane spanned by $(n_t, \Omega_{\mathrm{gw}} h^2)$ by following the dynamics of the inflaton. 
 
As we can see from Figs.~4, 5, and 6 these models of inflation are all consistent with a region in the $(n_s,r)$ plane that has a negative slope which reaches $r=0$ when $n_s = 1$.  It is for this reason that the \emph{upper} limit to $n_s$ is crucial when attempting to place a \emph{lower} limit to the amplitude of the IGWB for these models. Therefore, as we can see in the Figures, the WMAP3 constraints are not as restrictive as one might have thought since the degeneracy between $r$ and $n_s$ in the CMB allows for a larger value for $r$ compensated by a larger value of $n_s$.  As commented in the previous Section, the WMAP3+ constraints remove much of this degeneracy so that the constraint contour is more vertical in the $(n_s,r)$ plane and hence much more restrictive for these models. 

\section{Conclusions}
Recent measurements of the scalar spectral index indicate that it may be less than unity.  This fact has caused a great deal of excitement given that it is believed that having $n_s <1$ implies a significant amplitude for the gravitational-wave background produced by inflation.  
In this paper we have investigated this claim by analyzing predictions derived from six classes of models of inflation.  We have also extended the analysis to include not only the amplitude of the IGWB accessible to observations of the polarization of the CMB but also the IGWB accessible to direct observation.  

Our results can be divided into two different classes of inflationary potentials.  These classes are characterized by the curvature of the potential evaluated at the field value corresponding to CMB observations.  The curvature of the potential at a given field value is related to the sign of the slow-roll parameter $\eta$.  Models that have $\eta < 0$ (the inflaton is `falling off of a cliff') have decreasing $r$ as $n_s$ deviates further from unity.  Models that have $\eta > 0$ (the inflaton is `rolling down a bowl') have increasing $r$ as $n_s$ deviates further from unity.  This classification is directly related to the classification scheme presented in Ref.~\cite{Kinney:2000nc} in which inflationary models are said to be `large field', `small field' or `hybrid'.  In their classification scheme the sign of $\eta$ as well as its relation to $\epsilon$ is used to divide inflationary models.  However, in this paper we have emphasized how just the \emph{sign} of $\eta$ indicates how various constraints to $n_s$ affect the model's prediction for the IGWB. 

In attempting to set a lower limit to the expected IGWB accessible to direct observation these two different classes of models split up accordingly: with $\eta <0$ an increase in the lower limit to the amplitude of the IGWB is obtained with an improvement in the lower limit to $n_s$; with $\eta > 0$ an increase in the lower limit benefits from an improvement in the upper limit to $n_s$.  

In terms of the possibility of observing the IGWB directly, quoted sensitivities for a second generation BBO mission for a year long integration sets the lower limit to a detectible IGWB at $\Omega_{\mathrm{gw}} h^2 \gtrsim 10^{-17}$.  
 As can be seen in the figures, current constraints to $n_s$ for the six inflationary models considered here imply that a large region in parameter space for all six models will produce IGWB amplitudes within reach of BBO.  However, except for Higgs and PNGB inflation, there are regions of parameter space for which the IGWB amplitude can be arbitrarily small.  
 As the errors on $n_s$ shrink on both sides then, depending on the central value for $n_s$, each of the six models analyzed here may eventually predict a \emph{minimum} IGWB amplitude.  In particular, the Planck satellite is expected to attain 0.5\% in a determination of $n_s$ at a fiducial value $n_s = 0.957$ \cite{planck}.  This would then translate into a lower bound, $r \gtrsim 0.0046$, for Coleman-Weinberg inflation  which translates into $\Omega_{\mathrm{gw}}h^2 \gtrsim 1.61 \times 10^{-17}$ for direct observation. 

\begin{figure}[h!]
\centerline{\epsfig{file=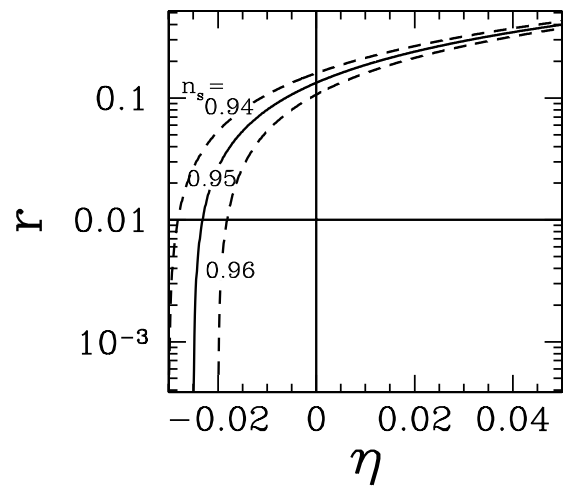,height=15pc,angle=0}}
\caption{An upper limit to $r$ along with a measurement of $n_s$ will tell us information on the curvature of the inflaton potential.  If we find that $r \lesssim 0.1$ and $0.94 \lesssim n_s \lesssim 0.96$ then we may conclue, within the context of single-field slow-roll inflation, that $\eta < 0$ which implies $V'' <0$.  Although qualitative, this conclusion would have far reaching implications for inflationary model building.}
\end{figure}

Barring a detection of the IGWB in the CMB our discussion shows that even an upper limit to $r$ and a precise measurement of $n_s$ tells us useful information on the curvature of the inflaton potential.  From the Eqns.~(\ref{ns}) and (\ref{r}) we can write
\begin{equation}
r = \frac{8}{3}(1-n_s + 2\eta).
\end{equation}
In Fig.~7 we show curves in the $(\eta,r)$ plane for $0.94 \leq n_s \leq 0.96$.  From that figure, we can see that for $n_s$ in this range an upper limit of $r \lesssim 0.1$ implies that the potential has a \emph{negative} curvature (this trend can also be seen in Fig.~9 in Ref.~\cite{Peiris:2006ug}).  This qualitative conclusion would have important implications for inflationary model building.  

\acknowledgments

We thank W. Kinney, D. Baumann, and D. Grin for useful discussions.  We thank E. S. Phinney for pointing out the erroneous factor of 2 in our previous expression for the gravitational-wave transfer function.  This work was supported at Caltech by DoE DE-FG03-92-ER40701,
NASA NNG05GF69G, and the Gordon and Betty Moore Foundation, and at UC Irvine by NSF CAREER AST-0645427. 

\bibliography{bibliography}{}

\end{document}